\documentclass
[12pt]
{iopart}
\usepackage{epsfig}

\begin{document}
\title{Continuum and all-atom description of the energetics of graphene nanocones}
\author{Antonio \v{S}iber}

\address{Department of Theoretical Physics, Jo\v{z}ef Stefan Institute, SI-1000 Ljubljana, Slovenia}
\address{Institute of Physics, P.O. Box 304, 10001 Zagreb, Croatia}
\ead{asiber@ifs.hr}

\begin{abstract}
Energies of graphene nanocones with 1 to 5 pentagonal disclinations are studied on an atomically 
detailed level. The numerical results are interpreted in terms of three different contributions to 
the cone energy: the core disclination energy, the bending energy of the cone surface, and 
the ''line tension'' energy of the cone edge that is related to different coordination of 
carbon atoms situated at the edge. This continuum description allows for a construction 
of analytic expressions for the cone energetics and indicates different regimes of 
cone sizes in which cones with a particular number of disclinations are preferred 
energywise. An important result of the study is that the energetics of various types of cones 
profoundly depends upon whether the dangling carbon bonds at the cone basis are saturated 
by hydrogen atoms or not. This may be of use for explaining the differences in the yields of various 
cone types in different production processes.
\end{abstract}
\date{\today}
\maketitle

\section{Introduction}

Carbon (C) atoms are famous for their ability to form various hollow structures\cite{Terroreview}, 
the best known of them being fullerenes\cite{Kroto1} and carbon nanotubes\cite{IIjima}.  
Soon after the clear experimental detection of carbon nanotubes \cite{IIjima}, 
conical carbon structures 
have been observed in the process of quenching hot carbon vapour on the graphite 
substrate \cite{Sattler1, Sattler2}. These cones all had the same opening angle, the smallest 
one of the five that can be constructed from the honeycomb graphene structure (see below).
It was later demonstrated that the structures 
of the same type can also be obtained by other experimental procedures, such 
as pyrolysis of hydrocarbons in plasma torch \cite{Krishnan}. In this procedure, five 
different types of cone geometries were observed. The yields of conical structures 
in the prepared samples can be quite high ($\sim$ 20 \%) \cite{Krishnan}. 
This fact alone indicates the importance of such structures. In addition, conical graphene 
structures are also predicted to have very specific mechanical 
\cite{Crespi1} and electronic \cite{Crespi2} properties, and could possibly be used as 
nanoscopic electron-field emitters. Interestingly enough, 
graphite cones were found in natural samples \cite{Jaszczak} which additionally 
highlights the need to understand the occurrence of conical shapes made of carbon.

The polyformity that is characteristic of hollow carbon shapes has its 
parallel in the structures formed by viral proteins. These also form hollow 
shells (the so called capsids) that protect the viral genetic material 
(DNA or RNA). Icosahedral viruses have a structure that is the same as 
the one exhibited by icosahedral fullerenes \cite{Sibvir}. The same proteins that form 
icosahedral shells may, under different conditions, also form 
tubular structures, similar in structure to carbon nanotubes. There seems 
to be a parallel for conical structures also - the core capsids of 
HIV viruses very much resemble (closed) graphene cone structures which 
was emphasised in Ref. \cite{Ganser}. The investigation 
of graphene cones may thus have an added value due to the parallelism 
of structures found in graphene shapes and viral capsids.

Unlike the fullerenes and carbon nanotubes, conical carbon structures are relatively 
poorly explored. Not much is known about the optimal conditions for their 
production. A complete understanding of growth of such structures is also lacking, 
which is not too surprising since the details of growth of the much more investigated 
fullerenes and carbon nanotubes are still missing \cite{Harris}. Although there 
have been some proposals that the formation of cones is most likely due 
to their ''designability'', i.e. the fact that they are a combinatorially favorable 
outcome of the growth \cite{MRSTreacy}, it is certainly important to first 
have a good understanding of and reliable estimates for the cone energies. 
Various contributions to the cone energetics are not completely self-evident, 
and it is not straightforward, for example, to estimate the importance of the 
curvature of the cone surface and its contribution to the total energy of the cone. 

In this article I shall examine the energetics of (single-walled) graphene cones. These cones 
contain multiple pentagonal defects. Cones with defective carbon 
rings whose number of sides is smaller than 5 are not considered in this study as 
these are shown to be less stable\cite{Ihara}. To understand 
various factors that influence the energy of the cones 
numerical, all-atom calculations of the cone energetics are performed. 
The numerical results are interpreted in terms of different energy 
contributions that are disentangled from the total energy balance. 
This enabled a construction of reliable {\em analytical} expressions 
for the cone energetics. 

Section \ref{sec:sec1} discusses the details of the cone geometry 
and describes a particular construction scheme of the cones that is 
used in this study.

Section \ref{sec:sec2} contains the details of the numerical calculations and 
their results. 

In Section \ref{sec:sec3} the numerical data obtained in Sec. \ref{sec:sec2} 
is interpreted in a quite general framework that enables a construction 
of analytic expressions for the cone energies. Four contributions 
to the cone energy, each of them having a different physical origin, are 
identified. 

Section \ref{sec:discuss} contains a discussion of the results and some 
propositions that may relate the results of the article with the 
experimental data on carbon cones.

\section{Construction of cones}
\label{sec:sec1}

There are five different classes of cones that can be constructed from the 
graphene lattice of carbon atoms. These differ in the total number of 
disclinations (i.e. pentagonal carbon rings) that they contain close 
to the cone apex, which is directly related to the cone opening angle, $\theta$.
The cones can be conceptually constructed by cutting out 60 degree wedges in the 
graphene plane, 
thereby creating a "cut-and-fold" cone ''origami'' that can be mathematically folded onto a 
three-dimensional pyramid-like object. Except for the cone with a single pentagonal 
disclination, this involves creasing along the lines connecting the 
neighboring pentagons. When the pyramid-like object is relaxed so that 
its energy is minimal, the shape acquires a cone-like appearance. Additional 
reconnections between the carbon atoms (e.g. between any two atoms that are on the 
cone edge and that could possibly form pentagons is they are sufficiently close 
\cite{ribbon}) are not allowed during the relaxation process.

Except for the 
cone containing a single pentagonal ring close to its apex, all other cones 
can be constructed in a multitude of ways, depending on the arrangement of 
the pentagonal disclinations (see e.g. Ref. \cite{JieHan}). 
These details are irrelevant for this study, as 
its main aim is to rationalize the cone energetics in the terms that 
belong to the realm of continuum physics. In each of the cone classes considered 
in the following, a 
fixed arrangement of the carbon pentagons is assumed. An isolated pentagon rule
\cite{IPR} was respected by all of the constructions so that the carbon pentagons do not share 
C-C bonds - each of the pentagons is completely surrounded by hexagons. 
The parameter that characterizes the cone is its size, i.e. the 
total number of carbon 
atoms that it contains. I shall also assume that the distances between the 
cone apex and its base, measured along the cone surface, are the same 
irrespective of the point on the base chosen. In other words, this means 
that the cones considered can be positioned on their base so that 
the cone apex projects exactly at the center of the base - the cones 
considered are right cones.

\begin{figure}[ht]
\centerline{
\epsfig {file=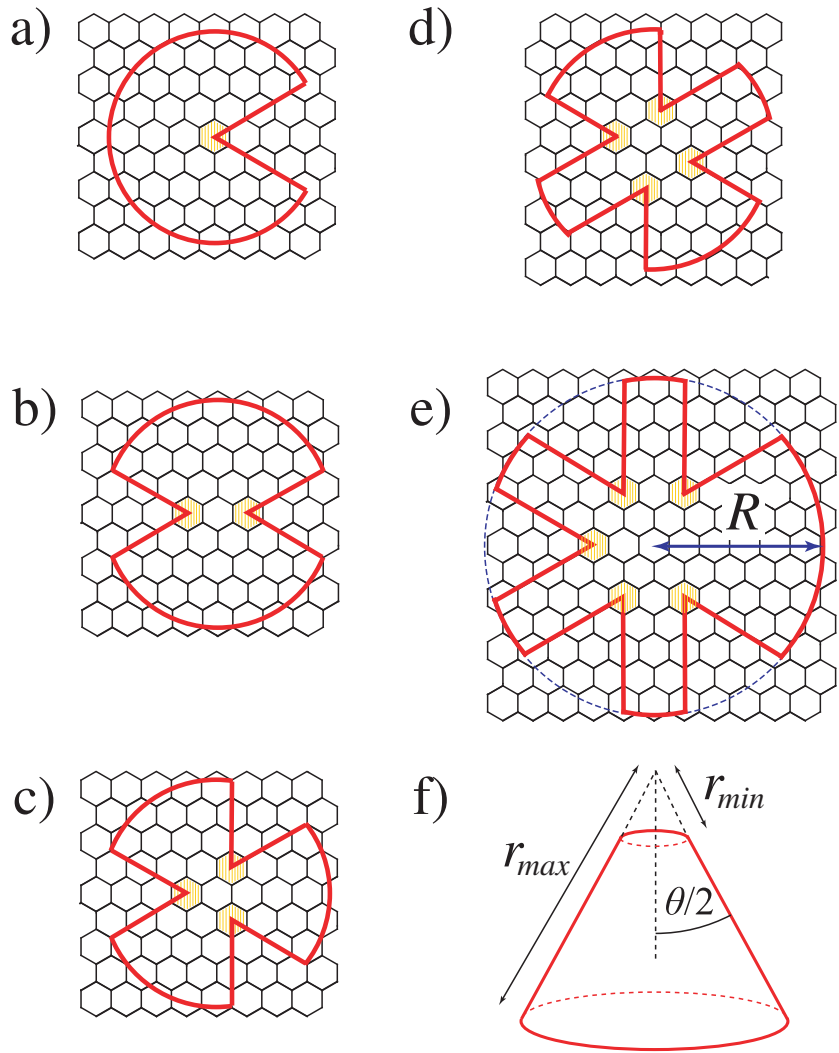,width=11cm}
}
\caption{Construction of cones by creation of disclinations (triangular wedges/cuts) 
in the graphene lattice. The part of the graphene plane bounded by the thick lines 
is folded into a cone. Panels a), b), c), d), and e) display the graphene 
cut-outs used to construct the cones with one, two, three, four, and five 
carbon pentagons, respectively. The size of the cones is determined by the 
radius ($R$) of the circle denoted by thin dashed line in panel e). Panel 
f) represents the cone and the quantities that define it (see text).}
\label{fig:figcone1}
\end{figure}

The five different cut-and-fold pieces of graphene plane are 
illustrated in Fig. \ref{fig:figcone1}. Similarly looking 
patterns were recently used for the construction of carbon nanotube 
caps, which necessarily have six pentagonal disclinations \cite{carbcap}. 
Note that the definition of 
the cone apex may be somewhat vague. In fact, it may be better 
to think of these structures as truncated cones or conical frusta. 
The smaller/top base of the conical frustum is then defined as 
the region surrounded the carbon pentagons (for graphene cone with 
a single disclination, the top base is the carbon pentagon itself). 
In the limit of very 
large cones, the distinction between the truncated cone 
and the regular cone becomes irrelevant since the radius of 
the bottom base of the cone is much larger than the radius of 
the cone top base.

It is easy to show that the opening 
angle of the idealized cone [$\theta$, see  Fig. \ref{fig:figcone1}f)] can be 
related to the total number of disclinations ($n$) as 
\begin{equation}
\theta_n = 2 \arcsin \left( 1-\frac{n}{6} \right).
\end{equation}
This produces a discrete sequence of possible opening angles, $\theta_n$ = 112.89, 
83.62, 60, 38.94, and 19.19 degrees, for $n=1,...,5$. It shall prove to be of use 
to consider a (planar) graphene disk as a special case of this construction. 
In this case, $n=0$, and $\theta_0=180$ degrees.

\section{Numerical results for the cone energies}
\label{sec:sec2}

The energetics of carbon-carbon bonding is simulated by using a second-generation 
reactive empirical bond-order potential by Brenner et al \cite{Brenpot} which 
belongs to the Abell-Tersoff class of bond-order potentials. This 
potential is known to properly account for the anisotropy of C-C bonding. It also 
approximately includes the many-body effects. After the mathematical folding 
of the piece of graphene plane was performed and the broken C-C bonds reestablished 
along the edges of the shape (this part of the procedure also forms the carbon 
pentagons), the resulting structure is used as an initial guess for the 
conjugate gradient procedure that optimizes its shape so as to minimize the 
total energy. The numerical procedure used for this purpose is described in 
details in Ref. \cite{CGrad}. Similar computational procedure 
(the construction of initial mathematically folded shapes and subsequent relaxation 
using the conjugate gradient method in combination with Brenner's potential) 
has also been used in Ref. \cite{Sibnano}.

\begin{figure*}[ht]
\centerline{
\epsfig {file=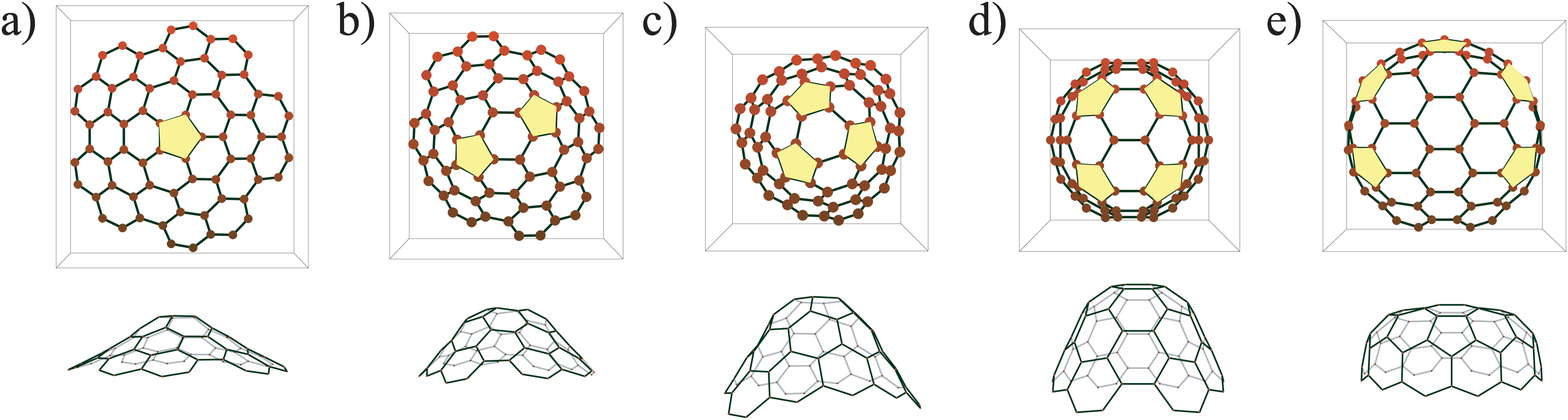,width=15cm}
}
\caption{Five different types of cones viewed from above (top row of images) 
and aside (bottom row of images). Panels a),b),c),d), and e) represent the 
$n=1$, $n=2$, $n=3$, $n=4$, and $n=5$ cones, respectively. The cones 
contain 70, 70, 69, 72, and 74 C atoms for $n=1,2,3,4$, and 5, respectively.}
\label{fig:figcone2}
\end{figure*}

Figure \ref{fig:figcone2} displays the top and side views of the five types of 
cones that have been relaxed to their minimum energy configuration. The cones 
chosen for display are relatively small so that the details 
of the shape, including the pentagons around the cone apex, can be easily 
discerned.

\begin{figure}[ht]
\centerline{
\epsfig {file=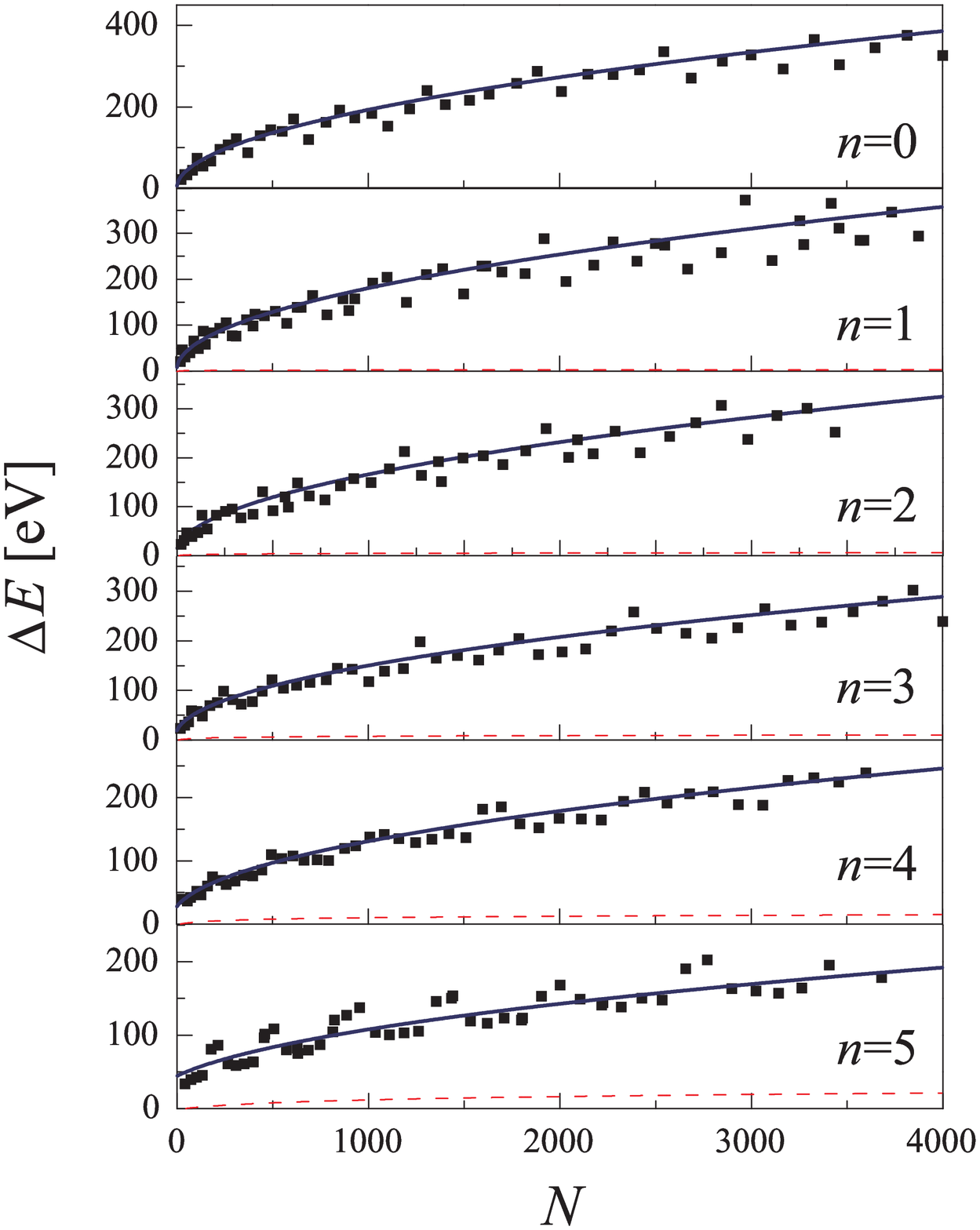,width=11cm}
}
\caption{Energies of carbon cones as a function of the total number of carbon atoms. 
In addition to results for $n=1$, $n=2$, $n=3$, $n=4$, and $n=5$ cones, the data for 
''$n=0$ cones'' i.e. graphene disks is also shown. The numerical results in these 
plots are denoted by full squares. Full thick lines are a result of analytical 
expression for the cone energy, Eq. (\ref{eq:analde}). The thin dashed lines denote 
the bending (curvature) contribution to the cone energy, Eq. (\ref{eq:conebend}).
}
\label{fig:figcone3}
\end{figure}

Figure \ref{fig:figcone3} displays the energies of the cones 
depending on the number of C atoms they contain. In these simulations, 
the radius of the cut-out shape [denoted by $R$ in Fig. \ref{fig:figcone1}f)] was 
varied. The energies displayed are in fact the {\em excess} energies ($\Delta E$), i.e. 
the extra energy that the cones have with respect to the energy that the 
same number of carbon atoms would have in the infinite graphene plane. 
Brenner's potential \cite{Brenpot} predicts that the binding energy per carbon atom 
is $\epsilon _b =$ -7.39494 eV \cite{Sibnano}, so that the energy of 
$I = N \epsilon _b$ was subtracted from the total energy of the cones in order to 
obtain the excess energy. The results for excess energies of graphene disks are also 
included in Fig. \ref{fig:figcone3}. Note the scatter of the data. It is not due to 
some numerical instability as shall be explained in the next section.

\section{General considerations of the cone energetics and interpretation of numerical data}
\label{sec:sec3}

To understand the numerical results obtained in previous section, it 
helps to isolate four different contributions to the energy of a cone. 
First, there is a part of the energy that is directly related to the 
number of diclinations in the cone. Each of the pentagonal disclination (defect) 
carries the so-called disclination core energy ($D$) \cite{Sibnano,Seung} 
that is related to change (reduction) of 
the local coordination of entities in the disclination. Second, the 
cones have a curved surface, so that there is a contribution to 
the total energy that is of the bending type ($B$) and directly related to nonvanishing 
curvature of the cone surface. For $n \neq 0$ this energy can be expressed as \cite{Sibnano}
\begin{equation}
B = \pi c_0 \frac{\cos^2 \theta_n}{\sin \theta_n} \ln \left(\frac{r_{max}}{r_{min}} \right),
\label{eq:conebend}
\end{equation}
where $c_0$ is the graphene bending rigidity \cite{Sibnano}, and $r_{min}$ and $r_{max}$ are 
the distances from the cone apex to the approximate top and bottom bases of the cone, 
respectively [see Fig. \ref{fig:figcone1}f)]. Equation \ref{eq:conebend} does not include 
the contribution related to bending of the shape surface in the top base, i.e. 
in the region where the disclinations accumulate. Assuming 
that the top base is nearly flat, or that the number of atoms that make 
it is much smaller from the total number of atoms in a whole conical shape, 
this part of the energy can be neglected. Third, 
there is an energy resulting from the coordination of the entities in a shape. This 
energy can be approximated as a sum of the energy that the same number of carbon atoms 
would have in the infinite graphene sheet (negative, binding contribution, $I$) and the 
cone-edge energy related to reduced coordination of atoms located along 
the line that defines the bottom cone base (positive contribution, $L$). The effect 
of the reduced number of neighbors for the atom at the cone edge could also be 
called the edge-tension energy and is obviously proportional to the length of the 
cone base. The total cone 
energy, $E$ is thus $E = n D + B + I + L$. 
Note that $B$, $I$, and $L$ depend on the size of the cone, i.e. on the total number 
of atoms, $N$, that make it, but in a very different functional manner. The excess 
energy of the cones is obviously given by
\begin{equation}
\Delta E = n D + B + L.
\label{eq:excess}
\end{equation}
The total area, $A$ of the conical shape can be approximately written as
\begin{equation}
A = \pi \sin \left( \frac{\theta_n}{2} \right) \left \{ r_{max}^2 - r_{min}^2 
\left [ 1 - \sin \left( \frac{\theta_n}{2} \right)  \right ] \right \},
\label{eq:area}
\end{equation}
so that the total number of atoms in the shape is
\begin{equation}
N = A / A_c,
\label{eq:number}
\end{equation}
where the area per atom in the infinite planar sheet of graphene 
is given by $A_c = r_{0}^2 \sqrt{3} / 2$. For the equilibrium nearest 
neighbor C-C distance in graphene, $r_0$, Brenner's potential 
predicts a value of \cite{Sibnano} $r_0$ = 1.4204 \AA.
The cone-edge energy depends on the total number of entities along 
the bottom basis of the cone. The total 
length of the bottom base edge is $2 r_{max} \sin (\theta_n / 2) \pi$, so that 
\begin{equation}
L = 2 \delta r_{max} \pi \sin \left( \frac{\theta_n}{2} \right),
\label{eq:tension}
\end{equation}
where $\delta$ is the average line tension energy per unit length of 
the cone base. Combining equations (\ref{eq:excess}), (\ref{eq:conebend}), 
and (\ref{eq:tension}) yields the analytical 
expression for the dependence of the excess energy on the $r_{max}$ parameter 
of the cone:
\begin{equation}
\Delta E = n D + \pi c_0 \frac{\cos^2 \theta_n}{\sin \theta_n} \ln \left(\frac{r_{max}}{r_{min}} \right) 
+ 2 \delta r_{max} \sin \left( \frac{\theta_n}{2} \right).
\label{eq:analde}
\end{equation}
The dependence of $r_{max}$ on $N$ can be obtained from equations (\ref{eq:area}) and (\ref{eq:number})
which yields
\begin{equation}
r_{max} = \sqrt{
\frac{N A_c}{\pi \sin \left( \frac{\theta_n}{2} \right)} + r_{min}^{2}\left[ 1-\sin \left( \frac{\theta_n}{2} \right) \right]
}.
\end{equation}
As mentioned earlier, the $r_{min}$ parameter is somewhat difficult to fix, but it can be estimated on 
the basis of constructions shown in Fig. \ref{fig:figcone1}. Values of 
$r_{min}=\{r_0 / \sin(\theta_1 / 2)$, $r_0 \sqrt{3} / \sin(\theta_2 / 2)$, 
$r_0 \sqrt{3} / \sin(\theta_3 / 2)$, $r_0 \sqrt{21} / [2 \sin(\theta_4 / 2)]$, 
$2 r_0 \sqrt{3} / \sin(\theta_5 / 2)$$\}$ were used 
for $n=\{ 1,2,3,4,5 \}$ constructions from Fig. \ref{fig:figcone1}, respectively. 
For $r_{max} \gg r_{min}$, the precise value of $r_{min}$ becomes irrelevant. It 
can be seen in Fig. \ref{fig:figcone3} that Eq. (\ref{eq:analde}) 
(full thick lines) nicely accounts for numerical data on cone energetics 
for all $n$ values. In these calculations, the core disclination energy and the graphene 
bending rigidity were set to $D=1.83$ eV and $c_0$=0.83 eV, respectively, in 
agreement with the values found in Ref. \cite{Sibnano}, and 
the best-fit value for $\delta$ was found to be $\delta = 1.30$ eV/\AA 
\hspace{0.7mm} (the same value was used for all $n$ values of the cones). This should 
be the edge energy lost per unit length of the edge due to the reduced coordination 
of the carbon atoms situated on the cone edge. The atoms at the edge can have 
one, two, but also three nearest neighbors (see cones in \ref{fig:figcone2}). 
The energy that is lost due to a missing neighbor should be about 
$2.46$ eV ($|\epsilon_b|/3$). The exact value of $\delta$ depends on the edge 
of a particular cone considered, i.e. 
the number of singly, doubly, and triply coordinated carbon atoms and their distances. 
These characteristics depend both on the construction of the cone and on 
its total size ($r_{max}$). This explains the reason for scatter of the data 
in Fig. \ref{fig:figcone3} - cones of different sizes generally have differently 
coordinated edges, so it may happen that in a cone of a particular size  
all of the edge atoms are doubly or triply coordinated, which is energetically favorable 
case (this is the case for all cones shown in Fig. \ref{fig:figcone3}). 
Some other sizes may have edges that contain a lot of singly coordinated 
edge atoms. This is an interesting combinatorial problem that I shall not 
dwell too much upon, as the details of the shape that depend on the 
precise arrangement of atoms are not of interest to this article. Nevertheless, 
it should be noted that the value of $\delta$ obtained from the fits can 
be nicely explained on the basis of energetics of C-C bonding. An important 
objection that could be made here is that the effective Brenner's potential 
cannot be reliably used in situations when there are dangling bonds. In Ref. 
\cite{Crespigrowth} it was suggested that Tersoff's version of carbon-carbon 
bonding overestimates the excess dangling bond energy. Complex non-pairwise-additive 
effects may be also expected in cases when there are dangling bonds 
and these could be beyond 
the reach of the potential model used in this study. Nevertheless, 
these effects can be expected only to renormalize the value of $\delta$, 
while the effective ''line-tension'' energy should still remain a useful 
concept even in that case. Even if the dangling bonds of C atoms at the cone edge 
were saturated by adsorption of hydrogen atoms \cite{Kobayashi,Balaban,Shenderova}, 
the reasoning in terms of the edge (line) tension energy would still be 
useful, only the value of $\delta$ would differ (see below). It is also appropriate 
to note here that the calculations that treated the dangling bonds in the 
same fashion as described above have been used in a molecular dynamics 
scheme to simulate the dynamics of carbon ''ribbons'' that may form 
cylinders (nanotubes) by saturating some of their dangling bonds \cite{ribbon}.

Looking at the data in Fig. \ref{fig:figcone3} it may seem that the 
edge tension energy dominates the cone energetics. This is evident 
both from the scatter of the data and from the apparent $\sqrt{N}$ behaviour 
of the excess energy. This is indeed the case except perhaps  
for $n$=5 cones, for which the bending contribution 
to the cone energy becomes important. This can be seen from the dashed 
lines in Fig. \ref{fig:figcone3} 
that indicate the bending contribution to the cone energy 
[Eq. (\ref{eq:conebend})]. This is in fact one of the important results of this 
study: the bending/curvature contribution to the cone energetics is 
secondary when compared to the dangling bond/edge energy, so the considerations 
that concentrate exclusively on this part of the cone energy in order to explain the 
occurrence of the cones with particular opening angle in fact deal with rather 
small contribution to the total cone energy \cite{Sattler1} (at least in cases 
when the bonds are not saturated by hydrogen atoms, see below).

\begin{figure}[ht]
\centerline{
\epsfig {file=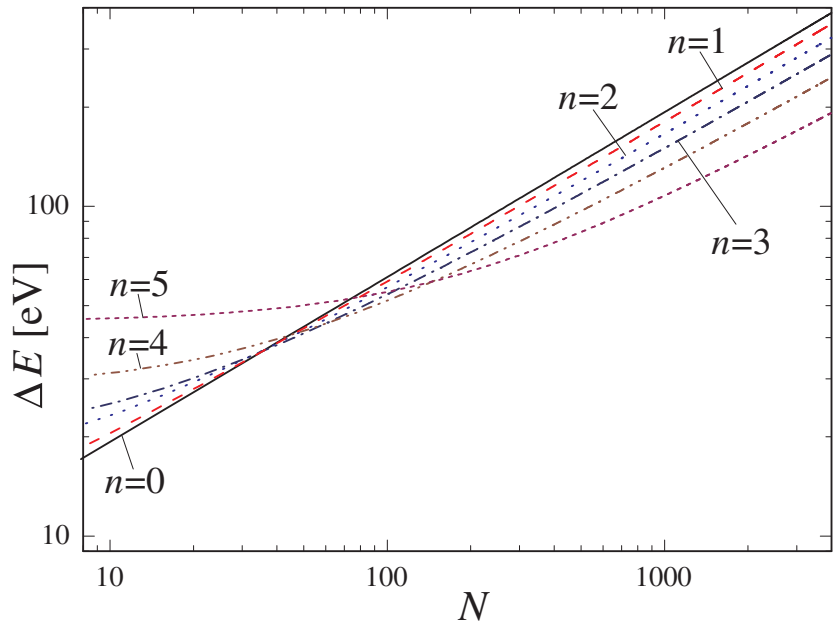,width=11cm}
}
\caption{Analytic (continuum) expressions for the cone energetics as a function 
of total number of atoms in the cones for the case when the dangling carbon 
bonds at the cone edge are not saturated ($\delta = 1.30$ eV/\AA ).
}
\label{fig:figcone4}
\end{figure}

It is instructive 
to compare the analytic expressions 
for the cone energetics in a single plot, as is done in Fig. \ref{fig:figcone4} 
(note the $\log$ scales on both axes and that the curve corresponding to 
graphene disks is a straight line). 
It can be seen that the graphene disks have the smallest excess energies for 
small number of C atoms ($N<50$). When the number of atoms in the structures reaches about 
$N=50$, the excess energies of all the cones except the $n=5$ cone, become smaller 
than the excess energy of the disk. At this point the radius of the disk is 
about 0.7 nm. When $N$ reaches about 150 atoms, the excess energy of $n=5$ cones 
becomes the smallest and remains so as $N$ increases. The transition region in 
the cone sizes that is observed in energetics of the cones smaller than about 
150 atoms is a consequence of the interplay of different contributions to the 
excess energy of the cone. After a critical size of the graphene disk ($N \approx 50$) the 
cone structures become energetically preferable, since the energy that is 
lost due to creation of disclinations is compensated by the reduced length of 
the cone edges in comparison with the length of the graphene disk. For 
large enough cones ($N>150$), this effect dominates 
the energetics of the structures considered, and $n=5$ cones thus have the lowest excess energies, 
although they contain five pentagonal disclinations and their surface
has the largest curvature.

It should be instructive to see how all these conclusions change if the dangling 
bonds were saturated by hydrogen atoms. This may be the case for carbon 
cones obtained by pyrolysis of hydrocarbons \cite{Krishnan}. The energy of 
C-C bond in graphene is\cite{Brenpot} 4.93 eV, while the energy of 
the C-H bond is\cite{Brenpot} 4.526 eV. Thus the energy that is lost when 
replacing the C-C with C-H bond is 0.404 eV. This is also the energy that 
is lost {\em per carbon atom} at the cone edge when one of its bonds has
been saturated by H atom. This is significantly smaller from 2.46 eV lost 
per carbon atom with a missing carbon neighbor. The line tension energy in this case 
is thus $\delta = 0.404/2.46 \times$1.30 eV/\AA$= 0.213$ eV/\AA , more 
than six times smaller than in the case of unsaturated bonds. 
\begin{figure}[ht]
\centerline{
\epsfig {file=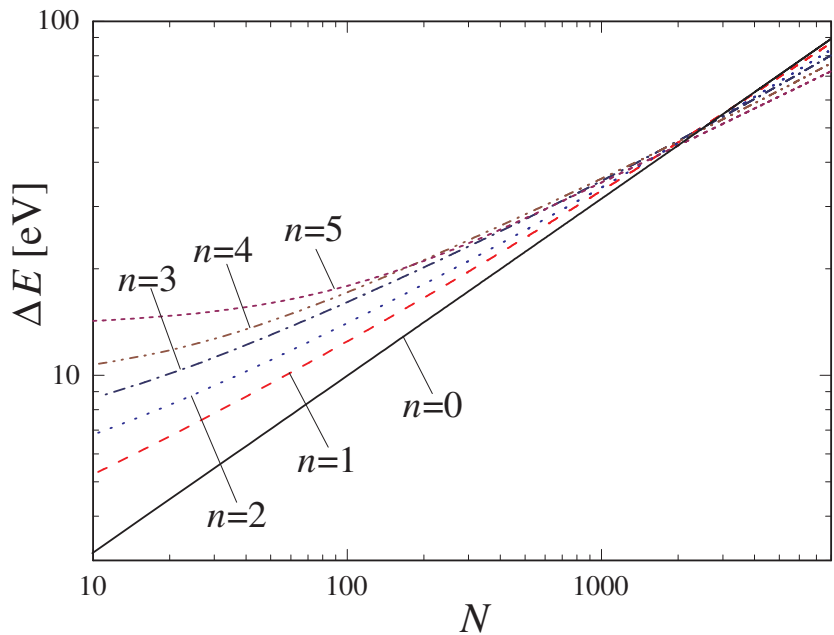,width=11cm}
}
\caption{Analytic (continuum) expressions for the cone energetics as a function 
of total number of atoms in the cones for the case when the dangling carbon 
bonds at the cone edge are saturated with hydrogen atoms ($\delta = 0.213$ eV/\AA ).
}
\label{fig:figcone5}
\end{figure}
Figure \ref{fig:figcone5} shows how the energetics of cones and disks whose edges 
are saturated by hydrogen atoms changes with the total number of carbon atoms in the 
structures. It can be seen that the graphene disks are the structures with the lowest 
excess energies up to total number of about $N=$2000 carbon atoms (compare this 
with $N=$50 in the unsaturated bond case). When $N>2000$, the cones have lower 
excess energies than graphene disks. Again, the $n=5$ cones are the most favourable 
in this respect. There is also an interesting crossover in cone energetics 
when $N<2000$ - note how the curve for $n=5$ crosses the $n=4$ and $n=3$ curves 
for $N \approx 170$ and $N \approx 700$, respectively. In this case, the 
bending contribution to the cone energy becomes important even for large 
cones. For example, for $n=5$ cone containing $\approx$9000 carbon atoms, 
the bending contribution to its energy is $B=27.1$ eV, while the edge energy 
is $L=39.1$ eV. 

Intriguingly, for the case of carbon cones whose edge atoms are saturated by 
hydrogen atoms, the flat disk structures have the lowest excess energies up 
to a much larger number of atoms than in the case of unsaturated, dangling 
carbon bonds at the cone edges. The crossover region at which the cones 
become preferred energywise is in this case positioned at significantly larger 
number of carbon atoms (2000 vs 150 for the dangling bonds case).

\section{Discussion}
\label{sec:discuss}

This study has demonstrated that graphene cones are superior to graphene 
disks with respect to their excess energy already for structures made of relatively 
small number of atoms ($N>50$ for the dangling bond case and $N>2000$ for 
the hydrogen saturated bonds at the cone edge). For structures containing 
very large number of atoms, $n=5$ cones always have the smallest 
excess energy. For the dangling bond case, of all the shapes considered in 
this study that contain more than $N=150$ atoms, the 
$n=5$ cones have the smallest excess energy. One could speculate that 
this may be one of the reasons for initial experimental observation of only 
$n=5$ structures \cite{Sattler1,Sattler2}. Later experiments have detected 
all possible cones \cite{Krishnan} and yields of cones of different $n$-type show 
a non-monotonic dependence on the number of disclinations, $n$. It would be 
tempting to assign this effect to the details of energetics of carbon 
cones in the cross-over region of sizes ($50<N<150$). This explanation would 
require that the stable seeds for the cone growth are nucleated in this interesting 
region of sizes. Of course, one should 
keep in mind that the growth of these structures is most likely governed 
by kinetics and entropy in addition to energetics. 

Intriguingly, it has been shown that the energetics importantly changes 
when one considers the cones whose edges are saturated by hydrogen 
atoms. In that case, the crossover region is positioned at much larger 
number of atoms, and the flat graphene disks are the lowest excess energy 
structures up to about 2000 carbon atoms. It is thus plausible that the 
yields of different cone structures will be different depending on 
whether the hydrogen atoms are present in the production process, assuming 
of course that they have an active role in stabilizing the 
carbon edges.

In summary, it has been demonstrated that the cone edge energy is the most important 
contribution to the total energy for cones that have more than about 150 
(2000) carbon atoms. The bending energy is significantly smaller from the dangling 
bond energy and becomes progressively less important for larger cones, 
since it scales with the total number of atoms, $N$, as $\log(N)$, while 
the dangling bond energy scales as $\sqrt{N}$. For cones whose edges 
are saturated by the hydrogen atoms, the consideration of the 
bending contribution to the total energy 
is necessary for the structures containing up to about 2000 carbon atoms.

\section{Acknowledgments}

This work has been supported by the Brain Gain Programme of the National Foundation 
for Science, Higher Education, and Technological Development of the Republic of Croatia 
under Award Number 02.03/25. and by the Ministry of Science, Education and Sports of 
Republic of Croatia through the Research Project No. 035-0352828-2837 (''Shapes and 
structures of nanoscale systems dictated by competition of energies'').

\end{document}